# Electrical conductivity measured in atomic carbon chains

O. Cretu[1], A. R. Botello-Mendez[2], I. Janowska[3], C. Pham-Huu[3],

J.-C. Charlier[2], and F. Banhart[1*]

[1] Institut de Physique et Chimie des Matériaux, Université de Strasbourg, UMR 7504 CNRS, 23 rue du Loess, 67034 Strasbourg, France

[2] Institute of Condensed Matter and Nanosciences, Université catholique de Louvain, Chemin des étoiles 8, 1348 Louvain-la-Neuve, Belgium

[3] Laboratoire des Matériaux, Surfaces et Procédés pour la Catalyse, UMR 7515 CNRS, 25 rue Becquerel, 67087 Strasbourg, France

* e-mail: florian.banhart@ipcms.unistra.fr

**Abstract**

The first electrical conductivity measurements of monoatomic carbon chains are reported in this study. The chains were obtained by unraveling carbon atoms from graphene ribbons while an electrical current flowed through the ribbon and, successively, through the chain. The formation of the chains was accompanied by a characteristic drop in the electrical conductivity. The conductivity of carbon chains was much lower than previously predicted for ideal chains. First-principles calculations using both density functional and many-body perturbation theory show that strain in the chains determines the conductivity in a decisive way. Indeed, carbon chains are always under varying non-zero strain that transforms its atomic structure from *cumulene* to *polyyne* configuration, thus inducing a tunable band gap. The modified electronic structure and the characteristics of the contact to the graphitic periphery explain the low conductivity of the locally constrained carbon chain.

Reducing carbon materials from three to two dimensions has revealed a new world of physical phenomena and technological applications in the past years. However, graphene as the two-dimensional allotropic form of carbon is not necessarily the lowest limit of spatial confinement. Narrow strips of graphene or small carbon nanotubes are already quasi one-dimensional (1D) systems exhibiting 1D signatures. However, linear chains of atoms are the thinnest imaginable and only real 1D allotropic form of carbon[1]. There had been many speculations about the existence and stability of *pure* atomic carbon chains[2] and about their properties until these 1D objects were finally observed in the electron microscope[3-6]. Indications for carbon chains have also been given, e.g., in the field emission characteristics of carbon nanotubes[7], in the electrical properties of a breaking nanotube[8], or during the coalescence of heat-treated carbon nanotubes[9]. However, these measurements could not yet be considered as direct evidence for the existence of carbon chains.

Theoretical studies have predicted interesting properties for pure atomic carbon chains. Indeed, an ideal carbon chain, also called *cumulene*, is characterized by σ-bonds along the axis (e.g. $s$-$p_x$ orbitals), and two decoupled π-bonds per atom from the perpendicular $p_y$ and $p_z$ orbitals. Similarly to higher-dimension carbon allotropes, both σ- and π-bonds in carbon chains are predicted to result in outstanding mechanical properties, e.g., a Young's modulus of more than one order of magnitude higher than diamond[10], as well as in a specific metallic behaviour. However, this 1D system is subject to Peierls instability, leading to a distorted lattice with bond length alternation (i.e., dimerization) and the appearance of a band gap[11]. This semiconducting behavior characterizes the *polyyne* configuration whose conductivity has been predicted to oscillate depending on an even or odd number of atoms in the chain[12, 13]. Furthermore, polyyne should exhibit spin-dependent transport properties[14]. Since the energetic stability difference between the perfect system (*cumulene*, with double bonds throughout the chain =C=C=C=C=C=) and the distorted system (*polyyne*, with alternating single and triple bonds -C≡C-C≡C-C≡) is rather small[15], both 1D systems could co-exist under appropriate experimental conditions. Therefore, carbon chains can be considered as ideal one-dimensional conductors and could find application as the thinnest possible wires for interconnecting ultimate nano-devices.

Notwithstanding the interest in carbon chains, their electrical properties have not been measured hitherto due to the difficulty of synthesizing and contacting these unstable objects. It has not even been tested if they are electrically conductive at all. Here, by using a scanning tunneling microscopy (STM) tip in a transmission electron microscopy (TEM) stage, both the synthesis and the electrical characterization of free-hanging atomic carbon chains are carried out *in-situ*. Surprisingly, their capability to carry an electrical current turned out to be much lower than expected and is discussed using state-of-the-art electronic-structure and quantum-transport calculations performed on specific configurations.

Few-layer graphene was obtained by the mechanical ablation of pencil lead, assisted by ultrasonification and followed by an acid/base purification[16]. The graphene flakes had a small number of layers and an average lateral size of 2 µm. Iron nanoparticles with an average size of 5 nm were deposited on the graphene flakes, leading to a catalytic cutting of the sheets. Pure iron and iron carbide particles as well as iron particles encapsulated in graphitic shells were obtained (details of the preparation procedure are described in the supplementary information). After dispersing the material in ethanol, a few drops of the suspension were deposited on a holey carbon TEM grid which had been previously cut in two halves. One of the half-grids was then mounted in a NanoFactory TEM sample holder[17, 18] that allows contacting exposed parts of the sample by precisely positioning an STM tip onto their surfaces. Gold tips prepared by electrochemical etching[19] were used as contacts. After establishing a contact between the Au tip and the half-grid, an electrical bias was applied between the sample and the tip, and the resulting current was measured. The in-situ experiments were carried out in a Jeol 2100F with $C_s$-corrected condenser, operated at 200 kV. The electrical measurements were monitored and recorded in real-time together with TEM images or videos of the contact region, giving a complete picture of the process. Under these operating conditions, the resolution of the microscope was around 0.2 nm while the current through the system was measured at the same time with a precision of the order of 1 nA. The temporal resolution for imaging and simultaneous electrical measurements was of the order of 0.1 s. An electron beam current density of approximately 50 A/cm$^2$ on the specimen was applied. The vacuum in the column was of the order 10$^{-5}$ Pa during the experiments. Images of the graphene flakes show a series of straight bands due to the interaction between the Fe nanoparticles and the graphene layers over which the Fe particles migrate[20]. The cutting of the flake by the moving Fe crystals leads to the formation of thin graphene nanoribbons standing off the edges (c.f. Fig. S1 of the supplementary material). These graphene ribbons were used as precursors for the formation of monoatomic carbon chains.

The *in-situ* synthesis of a carbon chain is summarized in Figure 1 (see supplementary Movie 1). The process started by contacting the edge of a few-layer graphene sheet with a Au tip. The first step was to ensure a good contact, which was either done by gradually increasing the voltage until the current through the circuit reached several 10$^{-4}$ A; typically at voltages around 1-1.5 V. Alternatively, the voltage was increased abruptly to higher values (2-3 V) while the current was limited to some $10^{-4}$ A in order to prevent the destruction of the graphene layers. Once such an electrical contact has been achieved, the voltage was decreased to around 1 V, and the tip was retracted slowly. As the part of the sample which is in contact with the tip starts to break from the larger flake, graphitic structures of less than 1 nm in width were formed between the two regions; this can be seen in Figure 1(a). Further separation of the tip from the sample resulted in the formation of a stretched graphenic structure (either flat or tube-like) and, eventually, in the unraveling of a monoatomic carbon chain from the graphenic layer (Fig. 1b). Under these conditions, the chain was found to be stable for a few seconds (Fig. 1b-e). The chain eventually broke and interrupted the electrical circuit (Fig. 1(f)). However, in most attempts the graphene ribbon detached completely from one of the electrodes without forming a chain.

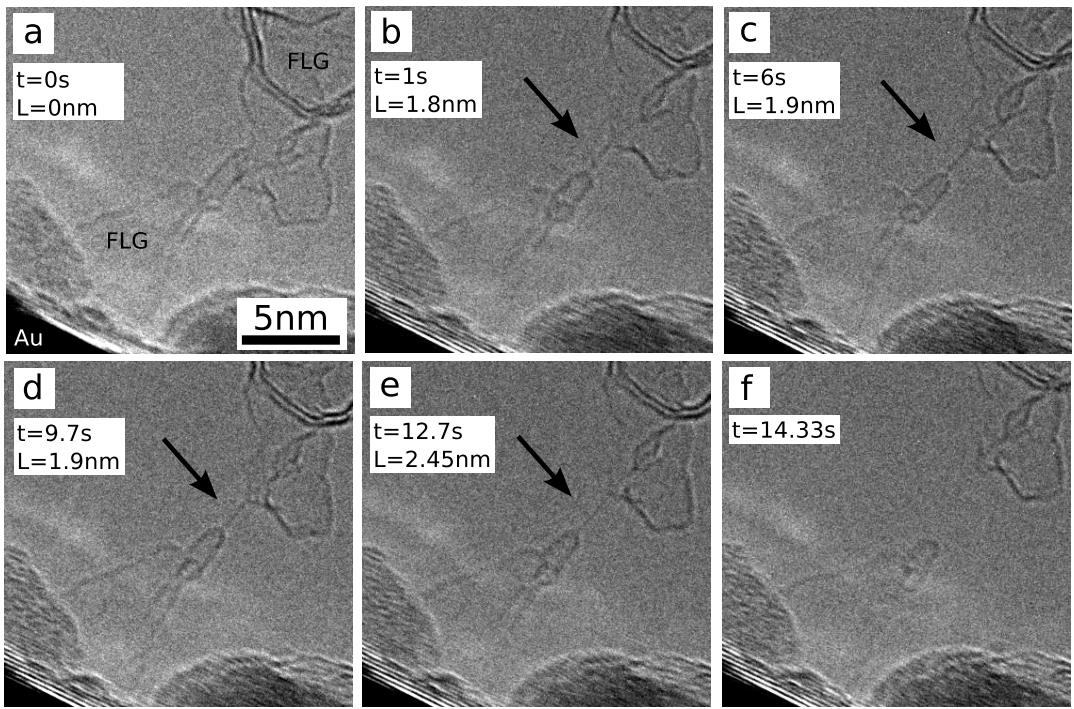

Fig. 1: *In-situ* synthesis of a monoatomic carbon chain. (a) A few-layer graphene nanoribbon (FLG) breaks and forms a carbon chain (arrowed) which is stable for a few seconds (b)-(e). The chain eventually breaks and disconnects the two FLG regions (f). The time scale as well as the measured length of the chain (in the projection onto the image plane) are indicated.

Each time when a chain unraveled from a graphenic structure, a characteristic drop of the electrical current was observed. Two examples are presented in Figures 2(a) and (b), more examples are shown in the supplementary information (fig. S3). The current drops abruptly from several $10^{-5}$ A (graphene ribbon) to values in the range $10^{-7}$ – $10^{-9}$ A (atomic chain). The low current persists for a few seconds and then drops to zero when the chain breaks and the circuit is eventually interrupted. By synchronizing the videos with the electrical recordings, it was found that the formation of the chains corresponds exactly to the moment where the current drops from micro- to nanoamperes. The low-current regime corresponds to the lifetime of the chains. This evolution was typical for all chains that were observed. The drop in current when the chain is formed proves that the linear contrast features are not projections of graphene ribbons in side-view.

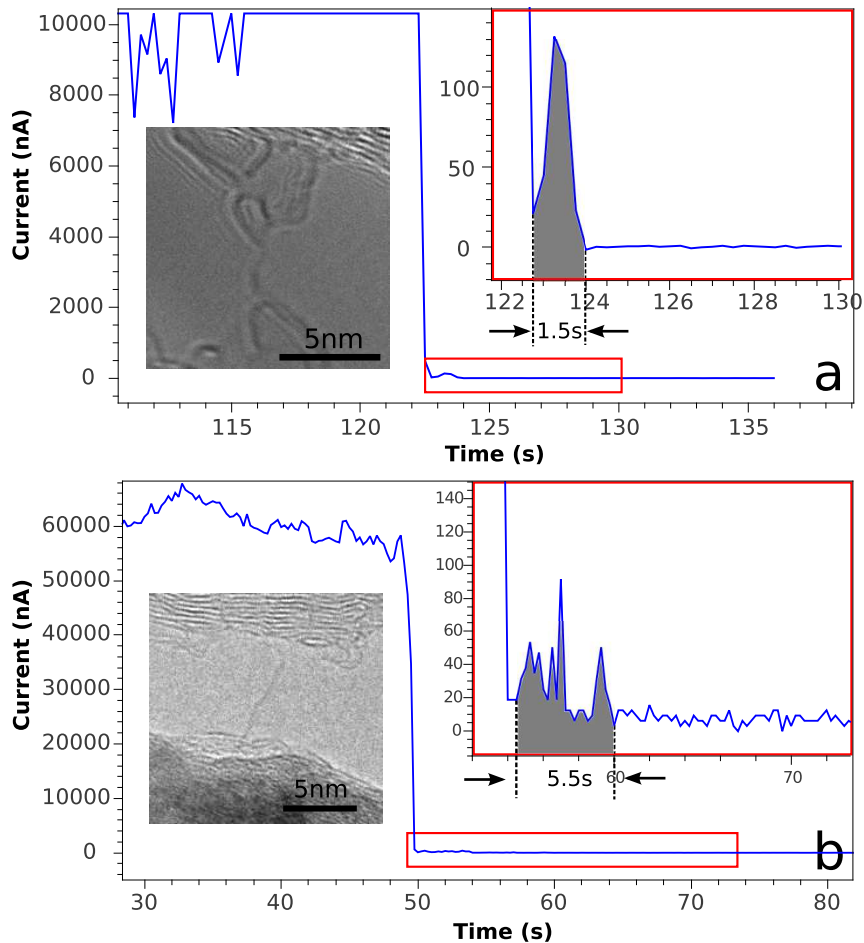

Fig. 2: Electrical current through two (a, b) monoatomic carbon chains at a constant bias of 1 V. The evolution of the current, monitored at the formation of the two chains, is shown (blue curve). The insets are zoom-ins of the areas in the red rectangles, in which the filled regions under the curve correspond to the presence of the chains. The image insets show the respective chains. The saturation of the current in the first part of (a) is the result of setting a ~10 µA current limit.

The short lifetime of the chains only allowed the recording of current-voltage characteristics in a few cases. Since the recording of an I-V curve was possible within 0.1 seconds, it was unaffected by the large fluctuations in conductivity that appear in Figure 2. Two I-V measurements for typical chains are presented in Figure 3 (more examples are shown in fig. S4 in the supplementary information). Both depict similar characteristics as the current changes by only 1-2 nA at a voltage variation of several hundred millivolts. Although the measurements of the low currents are affected by noise, both curves exhibit clear non-ohmic behaviour.

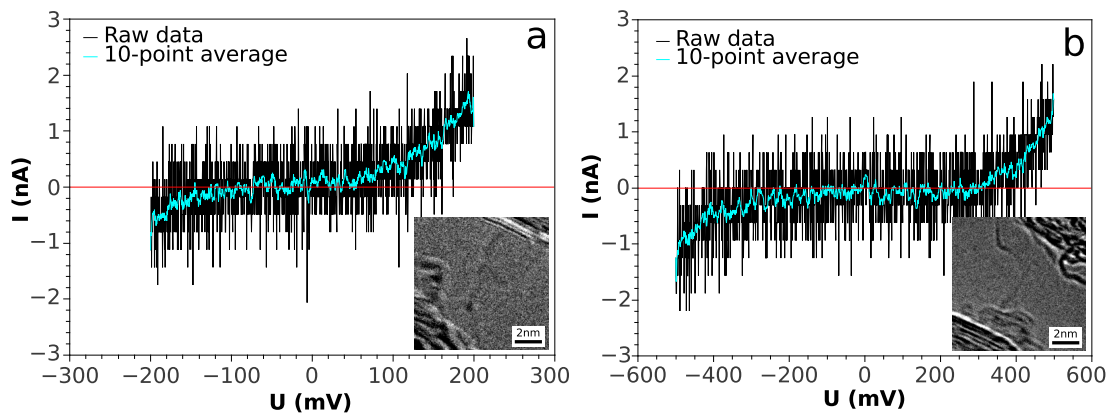

Fig. 3: I-V measurements on two monoatomic carbon chains. Images of the chains are displayed in the insets (the contrast in (a) is hardly visible due to the vibration of the chain). The discrete values of the noise are due to the sampling depth of the analog-digital conversion in the current recording.

The formation of the carbon chains can be explained by the unraveling of atoms from a graphene ribbon or a single-wall carbon nanotube[7]. In the present study, such a process is likely to occur due to the mechanical forces that are deliberately applied to the system and the locally high electric field between he electrodes. Several theoretical studies have already suggested the possibility of creating atomic carbon chains by inducing strain in graphene at room temperature[21-23] as well as at high temperature[24]. The role that defects in graphene[25] might play in this process has also been emphasized[4, 5]. By analyzing the evolution of the chains in the present experiments, a slow but ongoing unraveling of carbon atoms that increases the length of the chains is apparent. This might be responsible for the large fluctuations of current that was observed in most measurements (Fig. 2). A length increase of up to 36 % was seen which is much higher than the predicted ultimate strain of 12 %[26]. Hence, there is evidence that the structures under investigation continuously evolve. However, since different chains present quite different (though small) conductivities, strain in the linear arrangement should play an important role[15, 27].

In order to investigate the electronic and mechanical coupling in carbon chains, first-principles DFT and MBPT calculations (see Computing Techniques in the supplementary information) were performed to investigate both the electronic and structural properties in 1D carbon-based systems (Fig. 4). The *ab initio* electronic structure of a *cumulene* chain exhibits degenerate π bands crossing the Fermi energy (Fig. 4a), thus resulting in a quantum conductance of two quanta $G_0$ ($G_0=2e^2/h$) and a corresponding current of 15 µA at 0.1 eV bias (Fig. 4b). Within this highly symmetric configuration, the π orbitals are homogeneously distributed along the chain, forming equivalent bonds (Fig. 4e - top). As mentioned previously, such an ideal 1D system is subject to structural distortion. Indeed, Peierls' theorem predicts that a 1D system of equally spaced atomic sites with one electron per atom is unstable. A lattice distortion will cause the electrons to be at a lower energy than they would be in a perfect crystal, thus inducing the well-known Peierls dimerization in the chain. This new structure with bond-length alternation, becomes energetically favorable and the corresponding π orbitals are more localized on the shorter bond length, thus forming the triple bond (Fig. 4e - center). Due to the loss of symmetry, the *polyyne* chain exhibits an electronic band gap ($E_{Gap}$). The LDA(GGA) *ab initio* electronic structure of this chain presents an $E_{gap}$ of 0.28 (0.34) eV (Fig. 4c). However, although the DFT calculations usually give an efficient and accurate description of ground state properties (total energy, lattice constants, atomic structures, phonon spectra, …), the Kohn-Sham band structure systematically underestimates the band gap (often by more than 50%)[28]. In order to address excited-state properties and to calculate the band structure including electron-electron interactions (many body effects), MBPT calculations within the GW approximation for the self-energy are frequently used to provide corrected values for $E_{Gap}$ in better agreement with experiments[29]. The value of $E_{Gap}$ for the *polyyne* chain, estimated after the GW corrections to DFT/LDA is 0.41 eV (Fig.4.c – dotted red line). Such absence of electronic states is also clearly visible in the quantum conductance of the *polyyne* chain (Fig. 4d).

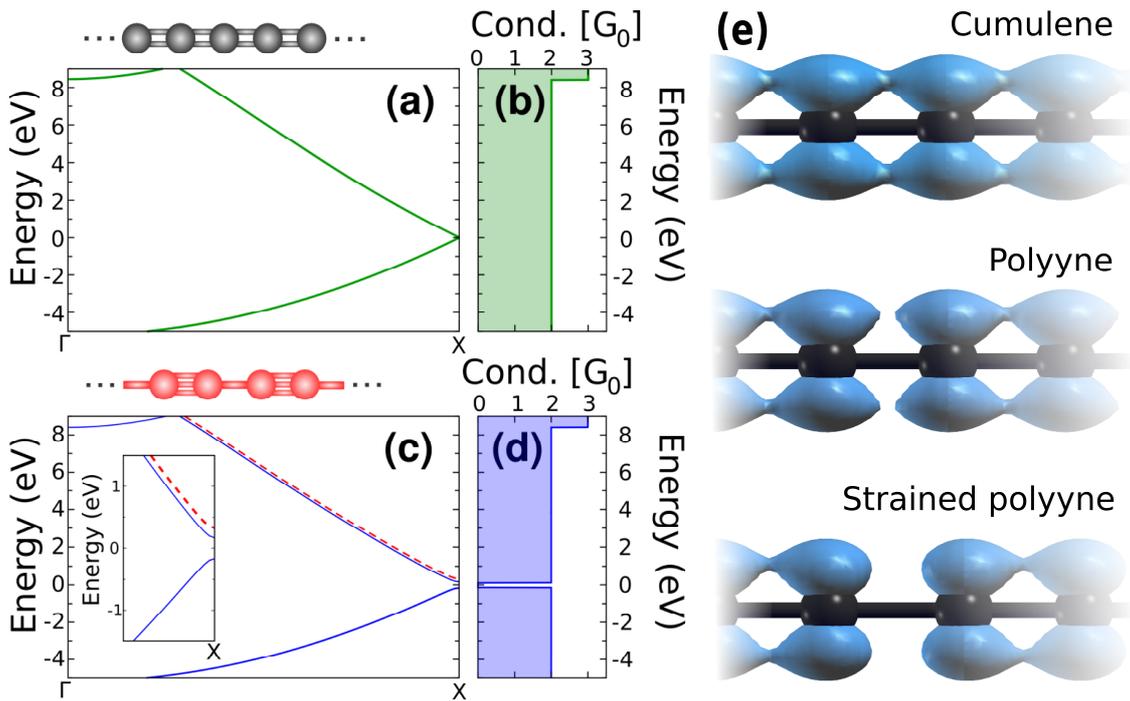

Fig. 4: *Ab initio* electronic structures and quantum conductances of *cumulene* (a, b) and *polyyne* (c, d) carbon chains. The inset of (c) illustrates the band structure of polyyne within DFT/LDA (blue curves) and GW (dotted red line). For more details, see Computing Techniques in the supplementary information. Electronic distribution ($\pi$ *orbitals in blue*) perpendicular to the chain (e) within the two atomic configurations (*cumulene* and *polyyne*) and under mechanical strain.

Figure 5a illustrates the strain-dependence of the electronic band gap (in both DFT and MBPT approaches) for an infinite *polyyne* chain. Note that 5% strain is enough to induce a 1 eV GW gap. In addition, as the strain increases, the dimerization (ΔL: difference of lengths between the triple and the single bonds in the *polyyne* chain) is more pronounced (Fig. 5b), thus stabilizing this specific configuration under strain[15]. Indeed, the electronic distribution perpendicular to the strained *polyyne* chain depicts that π orbitals are even more localized on the shorter bond length than for the pristine *polyyne* case (Fig. 4e – bottom). The large influence of the strain on the bandgap could be responisble for the different behaviour of the two chains in figs. 3 a and b that might be in different strain states. It should be pointed out that the bandgap cannot be measured in the present experiment because a three- or four-terminal setup would be needed. This would only be possible for a chain on a substrate that, in turn, would drastically modify the electrical properties and not allow us to characterize a pure carbon chain.

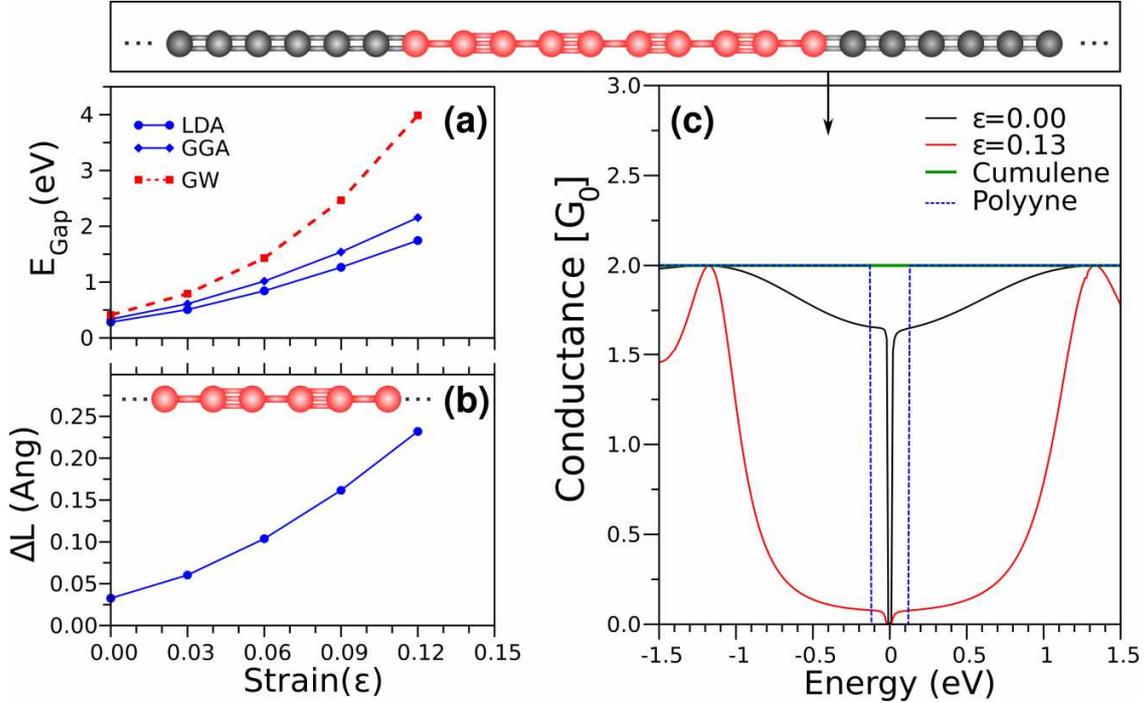

Fig. 5: Strain-dependence of the atomic structure, the electronic band gap, and the quantum conductance in carbon chains. (a) Electronic band gap versus strain in a *polyyne* chain within the approximations DFT/LDA (blue circles), DFT/GGA (blue diamonds), and GW (red squares). (b) Dimerization (ΔL : length of the longer bond minus length of the shorter one) versus strain in a *polyyne* chain. (c) First-principles quantum conductances of infinite ideal *cumulene* and *polyyne* chains (green and blue lines as in Fig. 4, respectively), and for a 10-atom *polyyne* segment embedded in a *cumulene* chain (model on top) with 0% strain (black curve) and 13% strain (red curve).

In order to estimate the transport in carbon chains, DFT calculations using open boundary conditions were used to predict the quantum conductance throughout a linear carbon chain[30]. The quantum conductance of the *cumulene* chain is characterized by two channels due to the two conduction and two valence bands that meet at the edge of the Brillouin zone (BZ), and thus gives a conductance of $2G_0$ (Fig. 4 a-b). Similarly, the polyyne chain presents a conductance of $2G_0$, except close to the charge neutrality point (Fig. 4 c-d). Due to the dimerization process inducing two different bond lengths, the loss of symmetry between the carbon atoms lifts the degeneracy between the conduction and the valence bands at the corner of the BZ (see dotted line in the inset in Fig. 4 c). A 1D system constituted by a *polyyne* segment embedded into two perfectly conducting semi-infinite *cumulene* chains (i.e. 10 carbon atoms as depicted in Fig. 5) is used as a realistic atomic model of a polyyne chain with seamless contacts. Such a model is used to highlight the effect of strain that could appear locally in the carbon chain. Indeed, a local constraint should stabilize a *polyyne*–type configuration locally (finite-size segment), thus perturbing transport along the chain. First-principles calculations predict that the quantum conductance is drastically quenched when the chain is under stress (Fig. 5c, red line) when compared to the unconstrained situation (Fig. 5c, black line). Recall that the DFT conductance does not reflect the predicted GW-gap for an infinite *polyyne* chain under strain.

However, this conductance drop under strain only accounts for a one order-of-magnitude drop in the current i.e., ca. 0.1 µA a few hundred meV outside the gap, and not for a two- or even three orders-of-magnitude drop as observed experimentally. Nevertheless, although the strain of 13% imposed to the atomic model to illustrate the loss in conductance leads to a 2eV gap (equivalent to 8% strain according to our GW results), the conductance should be even lower if calculated using a self-consistent GW scheme with updated wavefunctions, a highly computer-demanding technique that has recently demonstrated its success[31] although only in small molecular systems.

An important factor determining the electrical transport through a carbon chain is the nature of the contacts between the chain and the electrodes. The carbon chains synthesized in the present study are obtained by breaking larger carbon structures which leads to carbon-carbon bonds at the ends of the chain, as previously investigated theoretically[32]. Furthermore, cases were observed where one end of the chain detached and re-connected repeatedly at one of the graphitic contacts (shown in Fig. S2 of the supplementary information). Here, the end of the chain jumps between different positions on the graphitic support, but the current remains below 200 nA throughout this process, similarly to the 'fixed' chains. From the theoretical viewpoint, an accurate treatment of the contacts is challenging because the properties of these systems are very sensitive to the exact geometry of the bond between the chain and the graphitic material[33, 14]. However, it can be anticipated that the conductance should drop even more drastically due to interference and reflections that are not considered in the transport calculations presented above. The large fluctuations of the conductivity during the measurements is thus most likely due to both changing characteristics of the contacts and changes in strain. A hypothetical four-terminal device would be necessary to separate these two contributions experimentally. Nevertheless, as has been shown most recently for graphene ribbons[34], two-terminal measurements of nano-objects, that are only suspended at the two end-contacts, give valuable information about their electrical conductivity.

The influence of both temperature and electron irradiation has also to be taken into account. Due to the non-uniform structure between the electrodes, the estimation of Ohmic heating is difficult (electron-beam heating is negligible). While we can assume a considerable heating effect in the graphene ribbons[35] before the formation of the chain, the current through the chain itself is low so that heating should occur on a moderate scale. Furthermore, the thermal conductivity of both graphene and carbon chains is supposed to be very high[36], so the heat should dissipate efficiently towards the bulky electrodes. While electron irradiation does not cause persistent electronic excitations in systems with conduction electrons, ballistic atom displacements may occur[37]. By assuming a displacement threshold of 14 eV[38-40], we calculated a displacement rate of approximately 0.01 s$^{-1}$ for each carbon atom under the present irradiation conditions. At a chain length of 10 atoms a displacement would occur every 10 seconds. This is only slightly longer than the observed lifetime of a chain, so we may assume that the observed breakage of the chain may, at least in some cases, be due to the displacement of a carbon atom by the electron beam. Hence, under non-irradiation conditions, the lifetime of the chains might be longer.

Another important aspect is the high electrical field between the contacts (up to 0.1-1 V/nm in the present experiments). The formation (unraveling) and annihilation of the chains could be influenced by the field[2,7]. As shown in figs. S2 (suppl. information) and video 2, repeated attaching and detaching of the chains from the contact happens in some cases and can be explained by the electrical field. Such a process indicates the feasibility of recently proposed mechanisms for a voltage-programmable bistable atomic-scale switch in logical nano-devices[41]. Such a field-driven mechanism can also explain the repeated on-off switching in fig. 2b (inset) and the sudden onset of the current at a bias of 400 mV in fig. S4. After the rupture of the chain at a contact, it can be restored by the electrical field between the contacts, making such a device quite robust despite the low stability of the chains.

The present study shows for the first time that atomic carbon chains exhibit electrical conductivity, albeit clearly lower than predicted for ideal chains. The low conductivity is explained in terms of local strain in the chain and also due to the contact with the electrodes. Indeed, while measuring the electrical current, atomic carbon chains are always inherently strained and, thus, exhibit a varying but always non-zero bandgap. Since both strain and contact characteristics change considerably during the measurements, the conductivity never reaches the predicted values for ideal chains. In addition, our simulations show that a unique *cumulene* or *polyyne* configuration might not exist due to the varying strain and, correspondingly, varying bonding state due to the stabilization of the dimerization process under stress. This result might have important consequences on the understanding of the nature and stability of carbynes and related carbon-based linear structures. Furthermore, the present study shows that the characteristics of the contact between carbon chains and their periphery is difficult to control on the necessary (i.e., atomic) level. Nevertheless, carbon chains remain interesting candidates for interconnects of sub-nanometer dimensions. Further applications could be found as 1D sensors since the electronic transport should be drastically affected by adsorbents (such as chemical or biological species that can be attached to the chains) that, besides changing the electronic structure, should also have considerable influence on the strain in the chain.

**Acknowledgements**

Fundings by the French *Agence Nationale de Recherche* (project NANOCONTACTS, NT09 507527) and by the F.R.S.-FNRS of Belgium are gratefully acknowledged. This research is directly connected to the ARC on « Graphene StressTronics » sponsored by the Communauté Française de Belgique. Computational resources were provided by the CISM of the Université catholique de Louvain.

**Addiditonal Information**

Supplementary information accompanies the paper.

# Supplementary Information

**Electrical conductivity measured in atomic carbon chains**

O. Cretu[1], A. R. Botello-Mendez[2], I. Janowska[3], C. Pham-Huu[3],

J.-C. Charlier[2], and F. Banhart[1]

[1] Institut de Physique et Chimie des Matériaux, Université de Strasbourg, UMR 7504 CNRS, 23 rue du Loess, 67034 Strasbourg, France

[2] Institute of Condensed Matter and Nanosciences, Université catholique de Louvain, Chemin des étoiles 8, 1348 Louvain-la-Neuve, Belgium

[3] Laboratoire des Matériaux, Surfaces et Procédés pour la Catalyse, UMR 7515 CNRS, 25 rue Becquerel, 67087 Strasbourg, France

**A. Preparation of the starting material**

Few-layer graphene (FLG) was obtained by mechanical ablation of pencil lead, assisted by ultrasonification and followed by an acid/base purification [I. Janowska et al., *Carbon* **50,** 3106 (2012)]. The FLG flakes had usually 5-8 layers and an average lateral size of 2.5 μm. High resolution electron microscopy and Raman spectroscopy confirmed a low concentration of basal defects ($I_D/I_G$ of FLG ≤ $I_D/I_G$ of graphite). XPS revealed a low oxygen amount (5%) within FLG without thermal annealing. Good crystallinity of the FLG was also deduced from the low electrical resistance of the flakes of 1.6 kΩ, while adsorbed species were still present on the graphene surface. The resistance measurements were done under ambient conditions by a two-point method through the bottom gold electrodes with 2.5 μm gaps where only weak side contact between graphene and the electrodes is possible. In order to measure the electrical resistance of well-defined nanosize graphene sheets under vacuum conditions, a Nanofactory system in a TEM was applied. The large lateral size of FLG was reduced by the catalytic cutting of graphene sheets with previously deposited iron particles. Iron nanoparticles with an average size of 5 nm were prepared from $Fe(NO_3)_3 \times 9H_2O$ dissolved in an ethanol dispersion of FLG, followed by reduction with $NaBH_4$ at 60°C for 2h, then separation and drying at (r.t. 100°C) from ethanol. The Fe/FLG hybrid was then heated rapidly to 800°C in hydrogen (50 ml/min) - argon (150ml/min) co-flow for 40 min resulting in cutting trenches in the FLG surface. According to the work of D. R. Strachan, A.T.C. Johnson et al. the surface etching with moving iron particles is related to methane ($CH_4$) formation from graphene and hydrogen used during the heating [S. S. Datta et al., *Nano Lett*. **8**, 1912 (2008)]. Our investigation revealed in most cases an encapsulation of iron with graphitic shells with the formation of iron carbide on the outer surface of the iron particle after cutting, which could be related to a low content of hydrogen. Detailed studies of the cutting mechanism and the crystallographic direction of the cutting are under preparation. Iron nanoparticles can serve as a counter electrode during TEM resistance measurements where the Fe-FLG nanocontact was found to solve a sample-electrode contact problem.

**B. Supplementary figures and measurements**

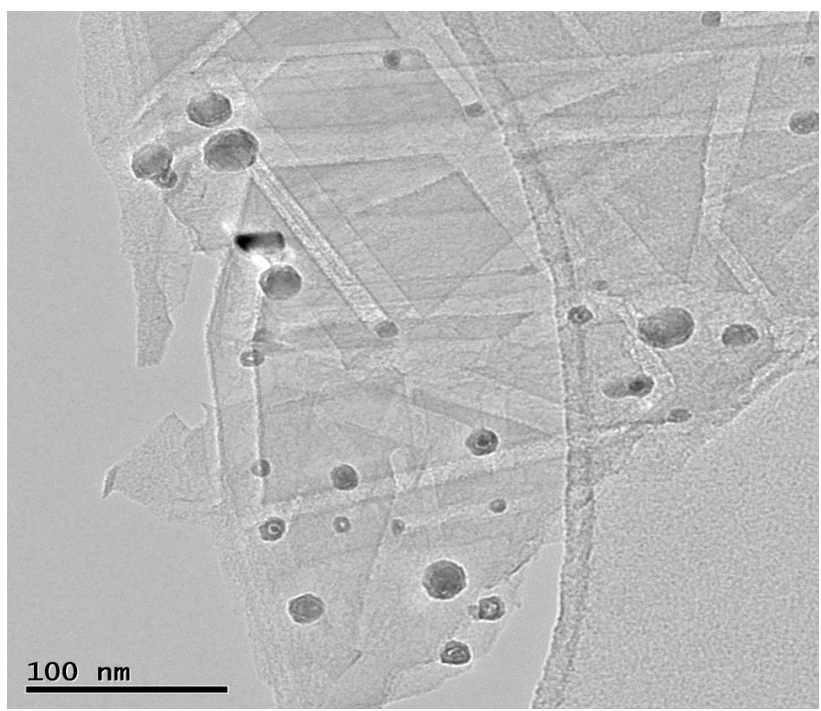

Fig. S1: TEM image of the starting few-layer graphene material, supported by an amorphous carbon membrane. Several linear traces due to the cutting of graphene ribbons by the moving Fe nanoparticles are visible.

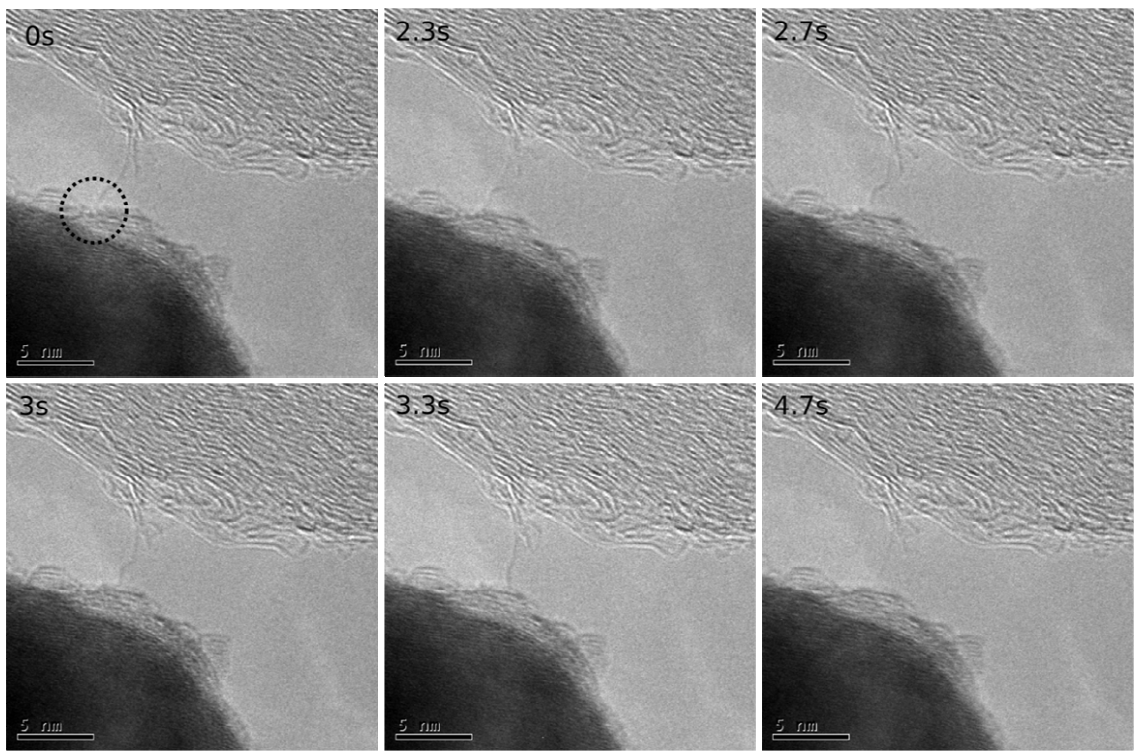

Fig. S2: TEM image sequence showing an unstable lower contact between the chain and the graphitic support. Repeated detaching and re-attaching leads to a sliding of the lower contact of a monoatomic carbon chain (encircled in the first image). See Movie 2.

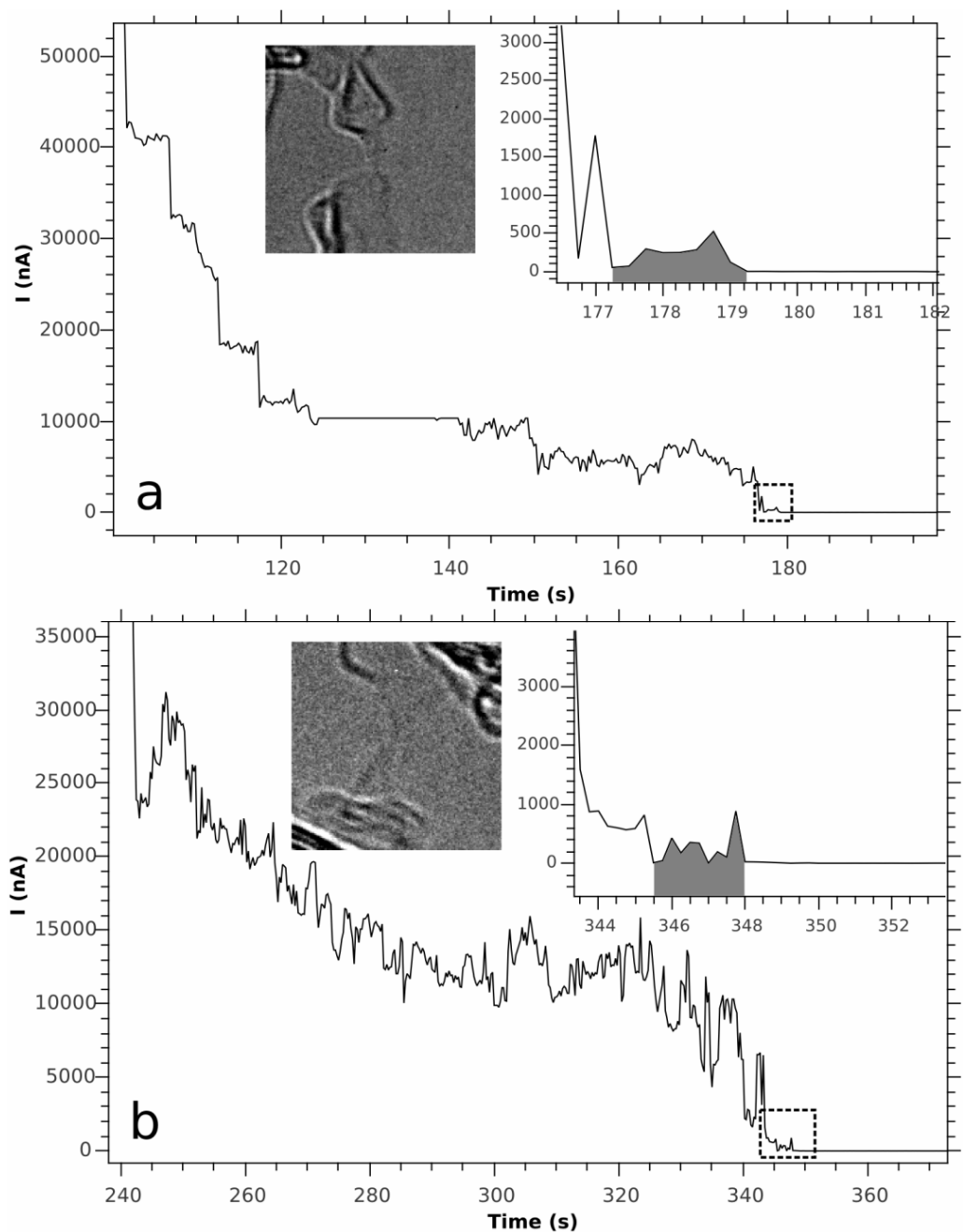

Fig. S3: Examples of current measurements through a graphene ribbon that transforms to a carbon chain. Two different cases are shown (a and b). The formation of the chains occurs at the final drop of the electrical current as shown in the inset diagrams (enlargements of the hatched boxes). See fig. 2 in the manuscript.

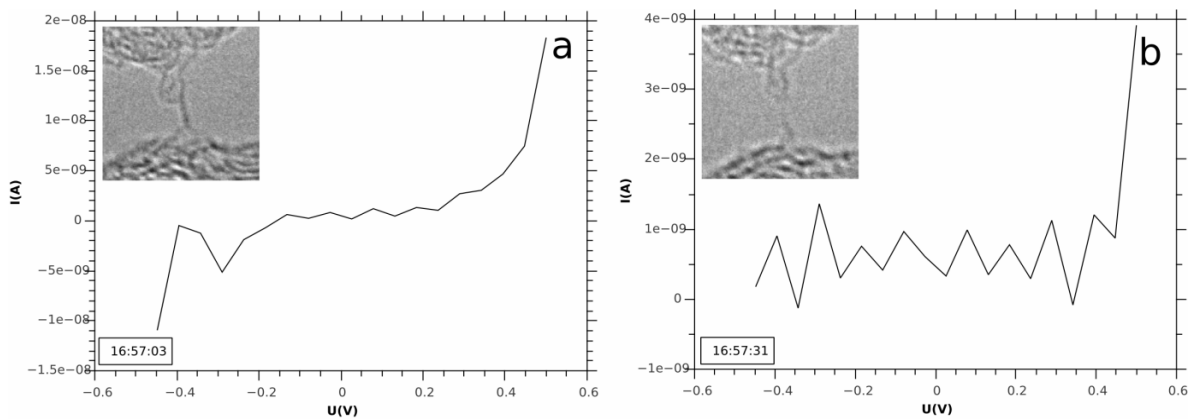

Fig. S4: Two current-voltage measurements through the same monoatomic carbon chain at different stages of evolution (the respective time is indicated on the bottom left display). In (a) the S-shaped curve shows the typical behaviour as in fig. 3 in the manuscript. In (b) the chain could already have been broken since the current remains zero in the voltage range +/- 400 mV. At higher positive bias, an onset of the current is seen, indicating a rebuilding of the chain in the electric field between the electrodes. For these measurements, a Keithley 2600 sourcemeter (Labview programming) has been used instead of the Nanofactory current-voltage measuring system.

**C. Computational Procedure**

Changes in the electronic structure of *polyyne* chains under strain were calculated using both Density Functional Theory (DFT) [P. Hohenberg & W. Kohn, *Phys Rev*. **136**, B864 (1964)] and Many Body Perturbation Theory (MBPT) within the GW approximation to the self-energy [L. Hedin, *Phys. Rev*. **139**, 796 (1965)], as implemented in the ABINIT package [X. Gonze et al., *Comput. Phys. Commun*. **180**, 2582 (2009)]. The exchange correlation is approximated by Ceperley-Alder (LDA) and Perdew-Burke-Ernzerhof (GGA). Norm conserving pseudo-potentials are used, and the wavefunctions are expanded on a plane wave basis set up to a kinetic energy cutoff of 36 Ha. The standard one-shot $G_0W_0$ approach was used to include quasiparticle corrections to the DFT-LDA eigenvalues. Quantum conductance calculations of segments of *polyyne* chains of different sizes embedded into *cumulene* were carried out using the Landauer formalism and the Green's function matching method [V. Meunier & B. Sumpter, *J. Chem. Phys.* **123**, 024705 (2005), E. Cruz-Silva et al. *ACS Nano* **3** 1913 (2009)] from a localized basis set DFT-GGA Hamiltonian [J. M. Soler et al., *J. Phys.-Condens. Mat*. **14**, 2745 (2002)].

**D. Supplementary videos**

Movie 1: In-situ synthesis of a monoatomic carbon chain. The video was recorded in real-time with a frame-rate of 3 fps.

Movie 2: Behaviour of a chain with an unstable contact as shown in Fig. S2. The video was recorded in real-time with a frame rate of 3 fps.